\documentclass[a4paper, oneside, twocolumn, notitlepage, 10pt]{extarticle_ecoc}
\usepackage{ecoc}
\pdfoutput=1
\newcommand{\thn}{\mathrm{th}}
\newcommand{\rin}{\mathrm{rin}}

\newcommand{\argmax}{\arg\!\max}
\newcommand{\argmin}{\arg\!\min}
\newcommand\notsotiny{\@setfontsize\notsotiny\@vipt\@viipt}

\begin{document}
\selectlanguage{english}    


\title{A New 5-bit/2D-symbol Modulation Format for Relative Intensity Noise-dominated IM-DD Systems}


\author{
    Felipe Villenas\textsuperscript{(1)}, Kaiquan Wu\textsuperscript{(1)},
    Yunus Can G\"{u}ltekin\textsuperscript{(1)}, Jamal Riani\textsuperscript{(2)}, and Alex Alvarado\textsuperscript{(1)}
}

\maketitle                  


\begin{strip}
    \begin{author_descr}

        \textsuperscript{(1)} Eindhoven University of Technology, 5600MB Eindhoven, The Netherlands,
        \textcolor{blue}{\uline{f.i.villenas.cortez@tue.nl}} 

        \textsuperscript{(2)} Marvell Technology Inc., Santa Clara, CA 95054, USA

    \end{author_descr}
\end{strip}

\renewcommand\footnotemark{}
\renewcommand\footnoterule{}

\begin{strip}
    \begin{ecoc_abstract}
        We propose a novel 5-bit/2D-symbol modulation format based on PAM-6 optimized for IM-DD systems dominated by relative intensity noise. The proposed modulation scheme improves SNR by 0.94 dB compared to conventional PAM-6 and achieves near-optimal BER performance. \textcopyright2025 The Author(s)
    \end{ecoc_abstract}
\end{strip}


\section{Introduction}
Intra-data center (DC) interconnects have experienced a continuous need for increasing data rates, especially driven by the exploding demand from artificial intelligence in recent years. The mainstream short-reach interconnect solution is optical communications using intensity modulation (IM) and direct detection (DD) with pulse amplitude modulation (PAM) formats. This approach enables cost-effective and low-power transceivers \cite{che2023modulation}.

Current commercial IM-DD systems for short-reach intra-DC interconnects have achieved up to 200 Gb/s/lane with PAM-4 \cite{800G_MSA_}. Achieving 400 Gb/s and beyond for next-generation systems is mostly hindered by two constraints: (i) the limited bandwidth of the electronic components in the transceivers \cite{pang2020200} and (ii) the relative intensity noise (RIN) from the transmitter laser \cite{szczerba20124}. The bandwidth limitation requires the use of higher-order PAM formats, such as PAM-6 \cite{hu2024467,uchiyama2024demonstration}. The most common PAM-6 generation scheme is to map every 5 bits to a point from a 32-point 2D constellation with a 6-point constituent alphabet. Then, the 1D components of the 2D symbols are transmitted over two consecutive uses of the IM-DD channel \cite{wei2018225,prinz2022comparison}. 

The use of PAM-6 requires a higher signal-to-noise ratio (SNR) than PAM-4 at the receiver in general, and thus transmit optical power must be increased. This increase in power generates RIN, which is a signal-dependent noise\cite{szczerba20124}. More specifically, the variance of RIN depends linearly on the square of the transmitted optical power, resulting in a varying noise distribution for each transmitted symbol. Therefore, investigating the effects of RIN for PAM-6 transmission is crucial for the design of next-generation IM-DD systems \cite{prinz2021pam}.

To the best of our knowledge, this paper presents the first optimization of a 32-point 2D constellation to improve the performance of PAM-6 transmission for IM-DD systems limited by RIN. First, the constellation is optimized to minimize the symbol error rate (SER). We determine the optimal 32-point subset of the square QAM-36 constellation that achieves the lowest SER. Then, we propose a novel binary labeling tailored to this optimized constellation. The bit error rate (BER) of this proposed labeled constellation approaches SER/5 asymptotically for large SNRs, mimicking Gray-labeled behavior, providing near-optimal performance\cite{agrell2004optimality}.


\section{Standard PAM-6 Transmission with RIN}


\begin{figure}[t]
    \centering
    \begin{adjustbox}{width=\columnwidth,trim={1.2cm 0 0.10cm 0}, clip}
        \definecolor{ashgrey}{rgb}{0.75, 0.75, 0.75}
\definecolor{antiquebrass}{rgb}{0.98, 0.81, 0.69}
\definecolor{brilliantlavender}{rgb}{0.96, 0.73, 1.0}

\newcommand{\OColor}{red}
\newcommand{\lw}{0.7pt}

\tikzstyle{block} = [draw, line width = 1pt, fill=white, rectangle, minimum height=30pt, rounded corners=0.1cm, text width=2.5em,align=center]

\tikzstyle{block_thin} = [draw, line width = 1pt, fill=white, rectangle, minimum height=20pt, rounded corners=0.1cm, text width=1.8em,align=center]

\tikzstyle{block_wide} = [draw, line width = 1pt, fill=white, rectangle, minimum height=30pt, rounded corners=0.1cm, text width=3.8em,align=center]

\tikzstyle{block_wide2} = [draw, line width = 1pt, fill=white, rectangle, minimum height=20pt, rounded corners=0.1cm, text width=3em,align=center]

\tikzstyle{block2} = [draw, line width = 1pt, fill=white, rectangle, minimum height=30pt, minimum width=30pt, rounded corners=0.1cm, text width=5em,align=center]

\tikzstyle{Cir} = [draw, circle,  minimum size=2.15em]

\begin{tikzpicture}[auto, node distance=1 cm,>=to,line width=0.5pt]

    \node [coordinate] (input) {};
    \node [block_wide, below = 1.0em of input, fill =black!10] (QAM_map) {QAM-32 Mod};
    \node [block_thin, below = 2.2em of QAM_map, fill = black!10] (P2S) {P/S};
    \node [block, right=2.5em of P2S, fill = blue!20] (DAC) {DAC};
    \node [block,right = 1.5em of DAC, fill = blue!20] (EO) {IM};
    \node [block_wide2,above = 1.5em of EO, fill=blue!20] (Laser) {Laser};
    \node[left=1.5em of Laser](RIN){RIN};
    \node[below=-0.4em of RIN](){($\sigma_{\rin}^2$)};

    \draw [solid,line width=\lw,opacity=1,\OColor] ($(EO.east)+(0.8,0.3)$) circle (3mm);
    \draw [solid,line width=\lw,opacity=1,\OColor] ($(EO.east)+(0.9,0.3)$) circle (3mm);
    \draw [solid,line width=\lw,opacity=1,\OColor] ($(EO.east)+(1.0,0.3)$) circle (3mm);
    \node[right=0.4em of EO, yshift=30pt,black!70] (oma){OMA};
    
    \node [block, right=1.8 cm of EO, fill = blue!20] (OE) {DD};
    \node[above=1.8em of OE, align=center](noise){AWGN ($\sigma_{\thn}^2$)};
    \node [block, right = 1.5em of  OE, fill=blue!20] (ADC) {ADC};
    \node [block_thin, right = 2.5em of ADC, fill = black!10] (S2P) {S/P};
    \node [block_wide, above = 2.2em of S2P, fill = black!10] (Decoding) {QAM-32 Demod};
    \node [above = 1.0em of Decoding] (output) {};

    \node[right=0.15cm of EO, yshift=-3.5em](ch_eq){Eq.~\eqref{eq:imdd_model}};
    \node[right=0.15cm of EO, yshift=-5.0em](ch_eq2){Eq.~\eqref{eq:ch_ele}};

    \node[above=8.75em of ch_eq](){IM-DD Optical Channel};
    
    \draw[draw,-stealth,rounded corners=4pt,dotted,gray,line width=\lw] ($(P2S.east)+(0.9em,-1.4em)$) |- (ch_eq);
    \draw[draw,-stealth,rounded corners=4pt,dotted,gray,line width=\lw] ($(S2P.west)+(-1.2em,-1.4em)$) |- (ch_eq);

    \draw[draw,-stealth,rounded corners=4pt,dotted,gray,line width=\lw] ($(QAM_map.south)+(-1.2em,-0.9em)$) -- ($(QAM_map.south)+(-2em,-0.9em)$) |- (ch_eq2);
    \draw[draw,-stealth,rounded corners=4pt,dotted,gray,line width=\lw] ($(Decoding.south)+(+1.2em,-1.2em)$) -- ($(Decoding.south)+(+2em,-1.2em)$) |- (ch_eq2);


    \draw [draw,-latex,line width=\lw] (input) -- node[above,text width=3.4em, yshift=0.2em,align=center]{Tx Bits}(QAM_map);

    \draw [draw,-latex,line width=\lw] (QAM_map) -- node[left,xshift=2pt,yshift=2pt]{$\boldsymbol{X}$}(P2S);
    
    
    \draw [draw,-latex,line width=\lw] (P2S) -- node[midway, below, xshift=-0.35em]{$X$}(DAC);
    \draw [draw,-latex,line width=\lw] (DAC) -- (EO);
    \draw [draw,-latex,line width=\lw, color=\OColor] (RIN) -- (Laser);
    \draw [draw,-latex,line width=\lw, color=\OColor] (Laser) -- (EO);
    \draw [draw,-latex,line width=\lw, color=\OColor] (EO) -- node[midway,below]{\textcolor{black}{Fiber}}(OE);
    \draw [draw,-latex,line width=\lw] (noise) -- (OE);
    \draw [draw,-latex,line width=\lw] (OE) -- (ADC);
    \draw [draw,-latex,line width=\lw] (ADC) --  node[midway,below,xshift=0.1em]{$Y$}(S2P);

    \draw [draw,-latex,line width=\lw] (S2P) -- node[right,xshift=-2pt,yshift=-2pt]{$\boldsymbol{Y}$}(Decoding);
    
    \draw [draw,-latex,line width=\lw] (Decoding) --  node[above,text width=3.4em,yshift=0.2em, align=center]{Rx Bits}(output);

    \draw [dotted,black!70,line width=\lw] (oma) -- ($(EO.east)+(0.5em,0)$);
    \draw [fill=black,black!70] ($(EO.east)+(0.5em,0)$) circle (1.5pt);
    

    \begin{pgfonlayer}{background}
        \draw[dashed,fill=blue!5,draw=blue!70,rounded corners=2pt] ($(DAC.west)+(-0.25,-0.85)$) rectangle ($(ADC.east)+(+0.15,2.2)$);
    \end{pgfonlayer}
    

    
    \node[above = 0.55cm of input, xshift=-2.5em] (elec_0){};
    \node[right = 1.7cm of elec_0] (opt_0){};
    
    \draw[draw,-latex,,line width=\lw] (elec_0) -- node[right,text width=4.0em, align = right]{\footnotesize Electrical}($(elec_0)+(1.7em,0)$);
    \draw[draw,-latex,\OColor,,line width=\lw] (opt_0) -- node[right,text width=3.2em, align = right]{\footnotesize \textcolor{black}{Optical}}($(opt_0)+(1.7em,0)$);
 
\end{tikzpicture}
    \end{adjustbox}
    \caption{IM-DD system under consideration based on PAM-6.}
    \label{fig:imdd_ch}
\end{figure}
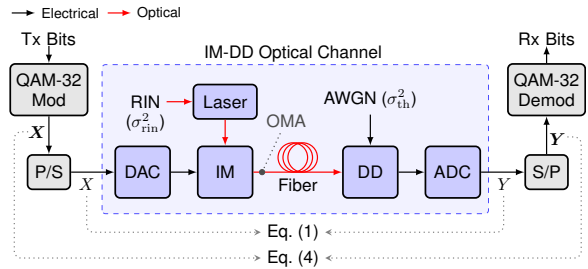

We consider the IM-DD system shown in Fig.~\ref{fig:imdd_ch}. For the optical transmission shown in the blue area, we use the same setup as in \cite{villenas2025ofc}, where there is no optical amplification present and a peak power constraint is imposed during IM\cite{che2023modulation}. The IM-DD optical link is modeled as a memoryless channel with additive noise via
\vskip -2mm
\begin{equation} \label{eq:imdd_model}
    Y = X + Z\sqrt{\sigma_{\thn}^2 + (X+\beta)^2\sigma_{\rin}^2},
\end{equation}
where \mbox{$X\in\mathcal{X}_{\mathrm{6}}=\{\pm1,\pm3,\pm5\}$} are PAM-6 symbols, $Z$ is a zero-mean and unit-variance Gaussian random variable, and $Y$ is the channel output. The constant \mbox{$\beta\geq |\min\{\mathcal{X}_{\mathrm{6}}\}|$} is a bias determined by the modulator dynamic extinction ratio \cite{che2023modulation} and is required to satisfy the non-negativity constraint for IM. At the fiber input, the transmitted optical power is measured through the optical modulation amplitude (OMA), defined as \mbox{$\mathrm{OMA}=\eta\left(\max\{\mathcal{X}_{\mathrm{6}}\}-\min\{\mathcal{X}_{\mathrm{6}}\}\right)$}, where $\eta$ is a parameter we use to vary the OMA.

The total noise variance is expressed as a function of the transmitted symbol $X$ as
\begin{equation} \label{eq:sigma}
    \sigma^2(X) = \sigma_{\thn}^2 + (X+\beta)^2\sigma_{\rin}^2.
\end{equation}
The contribution due to electrical noise is calculated from the receiver thermal noise equivalent power (NEP) as \mbox{$\sigma_{\thn}^2=(\mathrm{NEP}/\eta)^2\!\cdot\!B$}, where $B$ represents the noise electrical bandwidth. The laser noise variance is determined by the laser RIN parameter as \mbox{$\sigma_{\rin}^2 = 10^{\mathrm{RIN}/10}\!\cdot\!B$}\cite{szczerba20124}.

The standard generation of PAM-6 is achieved by decomposing a 32-point 2D constellation (QAM-32)\cite{chorchos2019pam,prinz2021pam}. Each 5-bit string is mapped to a QAM-32 symbol \mbox{$\boldsymbol{X}\in\mathcal{X}_{32}$}. The QAM-32 constellation is selected from the \textit{square} QAM-36, i.e., $\mathcal{X}_{32}\subset \mathcal{X}_{36}$, which is defined as
\begin{equation}
    \mathcal{X}_{36}=\{(a,b)\mid a,b\in\mathcal{X}_{\mathrm{6}}\},
\end{equation}
where $(a,b)$ is a row vector. 

After symbol mapping, a parallel-to-serial (P/S) block is used to separate the components of $\boldsymbol{X}$ as two consecutive PAM-6 symbols. Hence, for every QAM-32 symbol, the mapping to the PAM-6 symbols is given by $\boldsymbol{X}=(X_1,X_2)$.

After detection at the receiver in Fig.~\ref{fig:imdd_ch}, a serial-to-parallel (S/P) block is used to reconstruct the QAM-32 symbol as \mbox{$\boldsymbol{Y} = (Y_1,Y_2)$}. With $\boldsymbol{Y}$, the equivalent electrical channel can be modeled as a memoryless channel given by
\begin{equation} \label{eq:ch_ele}
    \boldsymbol{Y}=\boldsymbol{X}+\boldsymbol{Z},
\end{equation}
where $\boldsymbol{Z}=\left(\sigma(X_1)Z_1,\sigma(X_2)Z_2\right)$. Thus, we have that $\mathbb{E}[\boldsymbol{Z}^T\boldsymbol{Z}]=\mathrm{diag}\left(\sigma^2(X_1),\sigma^2(X_2)\right)$.

\begin{figure*}[!ht]
    \centering
    \input{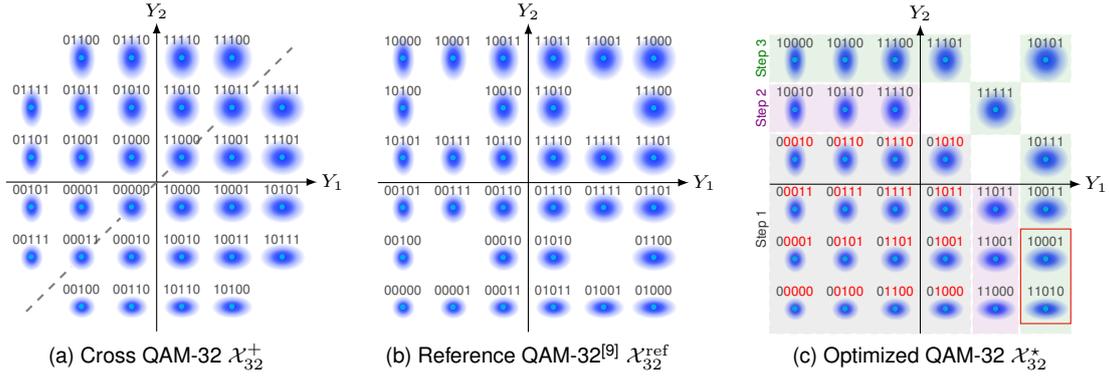}
    \caption{Comparison of QAM-32 constellations $\mathcal{X}_{32}$ employed in IM-DD based on PAM-6. The clouds indicate the variance of the noise distribution generated by RIN and AWGN. The bit labeling for each constellation is also shown.}
    \label{fig:constellations}
\end{figure*}

For equally-likely symbols, the optimal decision criteria for minimizing the SER is the maximum likelihood detection rule given by
\begin{equation} \label{eq:ml}
    \hat{\boldsymbol{x}} = \underset{\boldsymbol{x}'\in\mathcal{X}_{32}}{\argmax}\; p_{\boldsymbol{Y}|\boldsymbol{X}}(\boldsymbol{y}|\boldsymbol{x}'),
\end{equation}
where $\hat{\boldsymbol{x}}$ is the estimated received symbol, and $p_{\boldsymbol{Y}|\boldsymbol{X}}(\boldsymbol{y}|\boldsymbol{x})$ is the channel law in \eqref{eq:ch_ele}. After demodulation, the symbols and bits are recovered for SER and BER measurements.  

The \textit{cross} QAM-32 \cite{wesel2001constellation} is shown in Fig.~\ref{fig:constellations}(a). In this constellation, the corner points are removed with respect to a square QAM-36 $\mathcal{X}_{36}$, i.e., $\mathcal{X}_{32}^+=\mathcal{X}_{36} \setminus \{(\pm5,\pm 5)\}$. 
Due to RIN, the variance of the noise is symbol-dependent, and their components along the two axes are not identical. Furthermore, the distribution of the noise is symmetric with respect to the diagonal line $Y_2=Y_1$, as shown in Fig.~\ref{fig:constellations}(a). 
A different QAM-32 was constructed in \cite{prinz2021pam} by removing the middle point from each quadrant of the square QAM-36 as shown in Fig.~\ref{fig:constellations}(b), i.e., $\mathcal{X}_{32}^{\mathrm{ref}}=\mathcal{X}_{36} \setminus \{(\pm3,\pm 3)\}$.
This constellation was shown in \cite{prinz2021pam} to provide higher achievable rates and lower BERs for IM-DD systems for the AWGN channel, and thus it is expected to be suboptimal for the RIN channel. 


Additionally, based on eq.~(4) in \cite{renner2017experimental}, we define the SNR of the channel as\vspace{-1mm}
\begin{equation}\label{eq:snr}
    \mathrm{SNR}=\frac{\mathbb{E}[\|\boldsymbol{X}\|^2]}{\mathbb{E}[\sigma^2(X_1) + \sigma^2(X_2)]}.
\end{equation}
Furthermore, as $\mathrm{OMA}\!\to\!\infty$, the RIN becomes dominant and the SNR asymptotically results in\vspace{-1mm}
\begin{equation}\label{eq:snr_inf}
    \mathrm{SNR}_{\infty} \approx \frac{\mathbb{E}[\|\boldsymbol{X}\|^2]}{\mathbb{E}[(X_1+\beta)^2+(X_2+\beta)^2]\sigma_\rin^2},
\end{equation}
which only depends on the symbols $\boldsymbol{X}$ and the RIN variance $\sigma_\rin^2$.


\section{Optimizing QAM-32 Constellation}

The optimization we consider here is to minimize the $\mathrm{SER}$ over all the possible subsets $\mathcal{X}_{32}$ of $\mathcal{X}_{36}$, that is
\begin{equation} \label{eq:opt_problem}
    \mathcal{X}^\star_{32} = \underset{\mathcal{X}_{32}\subset \mathcal{X}_{36}}{\argmin}\;\mathrm{SER},
\end{equation}
where \mbox{$\mathrm{SER}=\mathrm{Pr}\{\hat{\boldsymbol{X}}\neq \boldsymbol{X}\}$}. 
The total number of possible constellations $\mathcal{X}_{32}$ is given by $\binom{36}{32}=58905$. To reduce the complexity of the optimization, we reduce the search space by exploiting the symmetry of the noise distribution with respect to the $Y_2=Y_1$ line (see Fig.~\ref{fig:constellations}(a)), which results in only $345$ combinations. We solved \eqref{eq:opt_problem} by exhaustively searching over all possible $\mathcal{X}_{32}$ that satisfy this symmetry. The optimal constellation found $\mathcal{X}_{32}^\star$ for the OMA region of interest is depicted in Fig.~\ref{fig:constellations}(c). Unlike the $\mathcal{X}_{32}^+$ and $\mathcal{X}_{32}^{\mathrm{ref}}$ constellations, $\mathcal{X}_{32}^\star$ is not symmetric around its origin. Instead, the 4 symbols removed from $\mathcal{X}_{36}$ are all located in the first quadrant where the total noise variance is larger due to RIN. 

Regarding bit labeling, we use a quasi-symmetric ultracomposite labeling for $\mathcal{X}_{32}^+$ as introduced in \cite{wesel2001constellation}. The reference constellation $\mathcal{X}_{32}^{\mathrm{ref}}$ is constructed to achieve a Gray labeling as shown in \cite{prinz2021pam}. Both labelings are shown in Fig.~\ref{fig:constellations}(a-b). On the other hand, our optimized constellation $\mathcal{X}_{32}^\star$ does not allow Gray labeling. To obtain a good labeling for $\mathcal{X}_{32}^\star$ with near-optimal performance\cite{agrell2004optimality}, we take a 3 step approach, as shown in Fig.~\ref{fig:constellations}(c). First, we start heuristically with a product of two PAM-4 binary reflected Gray code labelings for the bottom-left 16-point subset. Then, a new bit position is added (first bit position) which is going to be 0 for the previous subset, and 1 for the remaining 16 points. The six adjacent points are then labeled. Lastly, we tested all possible 10! combinations for the remaining 10 points, and selected the labeling with minimum average Hamming distance between the nearest neighbors. The resulting binary labeling is shown in Fig.~\ref{fig:constellations}(c). Most of the nearest neighbors have a Hamming distance of 1, except the two points shown in the red box, which have a Hamming distance of 3. 




\newcommand{\resultsFigs}{3} 

\ifthenelse{\equal{\resultsFigs}{2}}{
\newcommand{\xmin}{-8}
\newcommand{\xmax}{8}
\newcommand{\berLetter}{(b)}
\begin{figure*}[!ht]
    \centering
    \begin{subfigure}[b]{0.48\textwidth}
        \definecolor{my_green}{rgb}{0.55, 0.71, 0.0}    
\definecolor{my_purple}{rgb}{0.5, 0, 0.5}
\definecolor{azure}{rgb}{0, 0.5, 0.5}

\pgfmathsetmacro{\lw}{0.5}    
\pgfmathsetmacro{\lww}{0.3}    
\pgfmathsetmacro{\mksz}{1.2}    

\pgfplotstableread{FigTikZ/data_txt/SNR_RIN.txt}\datatableSNR

\newcommand{\blacksolid}{\raisebox{2pt}{\tikz{\draw[color=black,solid,line width=\lw,mark=o,mark options={solid}](0,0) -- (5mm,0);}}}   
\newcommand{\blacksdashed}{\raisebox{2pt}{\tikz{\draw[color=black,dashed,line width=\lw,mark=o,mark options={solid}](0,0) -- (5mm,0);}}}   
\newcommand{\blacksdotted}{\raisebox{2pt}{\tikz{\draw[color=black,dotted,line width=0.75pt,mark=o,mark options={solid}](0,0) -- (5mm,0);}}}   
\newcommand{\blackdiamond}{\raisebox{1pt}{\tikz{\node[draw,line width=\lww,scale=0.4,diamond,color=my_green,fill=white](){};}}}
\newcommand{\blackcircle}{\raisebox{1pt}{\tikz{\node[draw,line width=\lww,scale=0.5,circle,color=black,fill=white](){};}}}
\newcommand{\ksquare}{\raisebox{1pt}{\tikz{\node[draw,line width=1pt,scale=0.6,rectangle,color=red,fill=white](){};}}}

\newcommand{\pamfour}{orange!70!red}

\pgfplotscreateplotcyclelist{mylist}{
    {azure,line width=\lw,densely dotted,mark=diamond*,mark size=1.4 pt,mark options={solid,line width=\lww, fill=white}},
    {azure,line width=\lw,densely dashed,mark=triangle*,mark size=1.4 pt,mark options={solid,line width=\lww, fill=white}},
    {azure,line width=\lw,solid,mark=*,mark size=\mksz pt,mark options={solid,line width=\lww, fill=white}},
    {blue,line width=\lw,densely dotted,mark=diamond*,mark size=1.4 pt,mark options={solid,line width=\lww, fill=white}},
    {blue,line width=\lw,densely dashed,mark=triangle*,mark size=1.4 pt,mark options={solid,line width=\lww, fill=white}},
    {blue,line width=\lw,solid,mark=*,mark size=\mksz pt,mark options={solid,line width=\lww, fill=white}},
    {red,line width=\lw,densely dotted,mark=diamond*,mark size=1.4 pt,mark options={solid,line width=\lww, fill=white}},
    {red,line width=\lw,densely dashed,mark=triangle*,mark size=1.4 pt,mark options={solid,line width=\lww, fill=white}},
    {red,line width=\lw,solid,mark=*,mark size=\mksz pt,mark options={solid,line width=\lww, fill=white}}
}

\begin{tikzpicture}
    \begin{axis}[
    width=0.98\columnwidth,  
    height=2.5in,
    font=\footnotesize,
    xmin=\xmin, xmax=\xmax,
    ymin=13.5, ymax=28.5,
    xlabel={OMA [dBm]},
    ylabel={SNR [dB]},
    ytick pos=left,
    xtick pos=bottom,
    grid=both,
    grid style = {solid,lightgray!75},
    legend style = {legend pos=south east, font=\scriptsize, legend cell align=left, row sep=-0.5ex},
    ytick={14,16,...,28},
    xtick={-8,-6,...,8},
    cycle list name=mylist
    ]

        \addplot+[black] coordinates{(0,0) (1,1)};
        \addplot+[black] coordinates{(0,0) (1,1)};
        \addplot+[black] coordinates{(0,0) (1,1)};

        \addlegendentry{Regular}
        \addlegendentry{Reference}
        \addlegendentry{Optimized}
        
        \foreach \col in {1,2,...,9} {
            \addplot+[mark repeat=10,mark phase=0] table[x index=0,y index=\col] {\datatableSNR};
        }


        \node [font=\scriptsize,blue,rotate=5] at (axis cs:0, 21.65) {$\mathrm{RIN}\!=\!-141$};
        \node [font=\scriptsize,red,rotate=8] at (axis cs:1, 24.6) {$\mathrm{RIN}\!=\!-144$};
        \node [font=\scriptsize,azure,rotate=16] at (axis cs:-0.5, 26.85) {$\mathrm{RIN}\!=\!-147$};

        \draw[line width=\lww] (axis cs:6, 24) node[anchor=center, draw, ellipse, minimum width=6pt, minimum height=20pt] (circ){};
        
        \draw[draw, -latex, line width=\lww] ($(circ.west)+(1pt,-5.2pt)$) -- (axis cs:1,18);
        
    \end{axis}

    \begin{axis}[
        xshift=0.24\columnwidth,
        yshift=0.05in,
        width=0.43\columnwidth,
        height=1.25in,
        xmin=5.8, xmax=6.2,
        ymin=23.4, ymax=24.6,
        xtick={6},
        ytick={23.5,24,24.5},
        axis background/.style={fill=white},
        grid=none,
        grid style = {solid,lightgray!75},
        font=\footnotesize,
        cycle list name=mylist,
        xticklabel=\empty, 
    ]
        \foreach \col in {4,5,6} {
            \addplot+[red,mark repeat=3,mark phase=3] table[x index=0,y index=\col] {\datatableSNR};
        }


        \draw[<->, >=stealth, line width=\lww] (axis cs:5.83, 24.4326) -- (axis cs:5.83, 23.4967) node[midway, right,font=\scriptsize,yshift=+13pt,xshift=-2pt] {$_{\Delta_1=0.94\,\mathrm{dB}}$};
        \draw[<->, >=stealth, line width=\lww] (axis cs:5.86, 24.2055) -- (axis cs:5.86, 23.4967) node[midway, right,font=\scriptsize,xshift=-2pt] {$_{\Delta_2=0.71\,\mathrm{dB}}$};

    \end{axis}
    

    \node[font=\footnotesize] at (-0.3in, 1.85in) {(a)};
    
\end{tikzpicture}
    \end{subfigure}
    \begin{subfigure}[b]{0.48\textwidth}
        \definecolor{my_green}{rgb}{0.55, 0.71, 0.0}    
\definecolor{my_purple}{rgb}{0.5, 0, 0.5}
\definecolor{azure}{rgb}{0, 0.5, 0.5}

\pgfmathsetmacro{\lw}{0.5}    
\pgfmathsetmacro{\lww}{0.3}    
\pgfmathsetmacro{\mksz}{1.2}    

\pgfplotstableread{FigTikZ/data_txt/BER_RIN_144.txt}\datatable
\pgfplotstableread{FigTikZ/data_txt/BER_RIN_141.txt}\datatablee
\pgfplotstableread{FigTikZ/data_txt/BER_RIN_147.txt}\datatableee

\newcommand{\blacksolid}{\raisebox{2pt}{\tikz{\draw[color=black,solid,line width=\lw,mark=*,mark options={solid}](0,0) -- (3mm,0);}}}   
\newcommand{\blacksdashed}{\raisebox{2pt}{\tikz{\draw[color=black,densely dashed,line width=\lw,mark=o,mark options={solid}](0,0) -- (3mm,0);}}}   
\newcommand{\blacksdotted}{\raisebox{2pt}{\tikz{\draw[color=black,densely dotted,line width=\lw,mark=o,mark options={solid}](0,0) -- (3mm,0);}}}   

\newcommand{\blackdiamond}{\raisebox{1pt}{\tikz{\node[draw,line width=\lww,scale=0.4,diamond,color=my_green,fill=white](){};}}}
\newcommand{\blackcircle}{\raisebox{1pt}{\tikz{\node[draw,line width=\lww,scale=0.5,circle,color=black,fill=white](){};}}}
\newcommand{\ksquare}{\raisebox{1pt}{\tikz{\node[draw,line width=1pt,scale=0.6,rectangle,color=red,fill=white](){};}}}

\newcommand{\pamfour}{orange!70!red}

\pgfplotscreateplotcyclelist{mylist}{
    {line width=\lw,densely dotted,mark=diamond*,mark size=1.4 pt,mark options={solid,line width=\lww, fill=white}},
    {line width=\lw,densely dashed,mark=triangle*,mark size=1.4 pt,mark options={solid,line width=\lww, fill=white}},
    {line width=\lw,solid,mark=*,mark size=\mksz pt,mark options={solid,line width=\lww, fill=white}}
}

\begin{tikzpicture}
    \begin{semilogyaxis}[
    width=0.98\columnwidth,  
    height=2.5in,
    font=\footnotesize,
    xmin=\xmin, xmax=\xmax,
    ymin=1e-8, ymax=2e-1,
    xlabel={OMA [dBm]},
    ylabel={Bit Error Rate},
    y label style={yshift=-2pt},
    ytick pos=left,
    xtick pos=bottom,
    grid=both,
    grid style = {solid,lightgray!75},
    legend style = {legend pos=south west, font=\scriptsize, legend cell align=left, row sep=-0.5ex},
    ytickten={-8,-7,...,0},
    xtick={-10,-8,...,8},
    cycle list name=mylist,
    ]
        \addplot+[black] coordinates{(0,0) (1,1)};
        \addplot+[black] coordinates{(0,0) (1,1)};
        \addplot+[black] coordinates{(0,0) (1,1)};

        \addlegendentry{Regular}
        \addlegendentry{Reference}
        \addlegendentry{Optimized}
        
        \foreach \col in {1,2,3} {
            \addplot+[blue,mark repeat=5,mark phase=0] table[x index=0,y index=\col] {\datatablee};
        }

        \foreach \col in {1,2,3} {
            \addplot+[red,mark repeat=5,mark phase=0] table[x index=0,y index=\col] {\datatable};
        }

        \foreach \col in {1,2,3} {
            \addplot+[azure,mark repeat=2,mark phase=0] table[x index=0,y index=\col] {\datatableee};
        }




    \node [font=\scriptsize,blue] at (axis cs:5.5, 2.4e-3) {$\mathrm{RIN}\!=\!-141$};
    \node [font=\scriptsize,red] at (axis cs:5.5, 8e-5) {$\mathrm{RIN}\!=\!-144$};
    \node [font=\scriptsize,azure] at (axis cs:5.7, 2.5e-7) {$\mathrm{RIN}\!=\!-147$};

    \end{semilogyaxis}

    \node[font=\footnotesize] at (-0.4in, 1.85in) {\berLetter};
    
\end{tikzpicture}
    \end{subfigure}
    \caption{Performance comparison of the three QAM-32/PAM-6 constellations for (a) SNR, (b) SER, and (c) BER against OMA for different RIN values. The dotted lines correspond to the cross-shaped (regular) constellation, dashed lines to the reference constellation\cite{prinz2021pam}, and solid lines to the optimized constellation proposed in this work. All the RIN parameters in units of dB/Hz are depicted with the color of the corresponding curves.}
    \label{fig:SER}
\end{figure*}
}
{
\newcommand{\xmin}{-8}
\newcommand{\xmax}{8}
\newcommand{\berLetter}{(c)}

\begin{figure*}[!ht]
    \centering
    \input{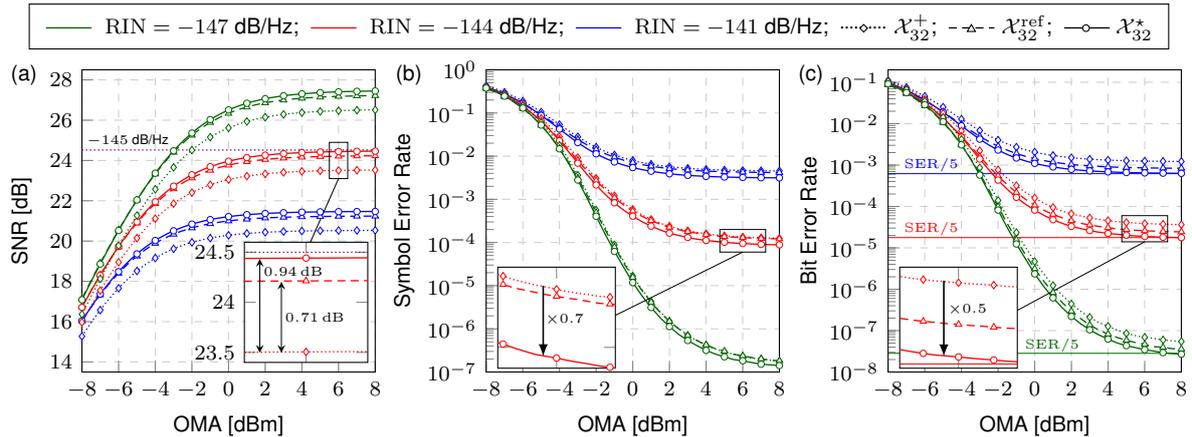}
    \caption{Performance comparison for the three constellations in Fig.~\ref{fig:constellations}: (a) SNR, (b) SER, and (c) BER against OMA for different RIN values. The dotted, dashed, and solid lines correspond to the cross, reference \cite{prinz2021pam}, and optimized QAM-32, respectively. The horizontal lines in (c) represent the approximated Gray label BER values for high OMA.}
    \label{fig:SER}
\end{figure*}
}

\section{Results}

We simulated the IM-DD system in Fig.~\ref{fig:imdd_ch} with the same parameters from \cite{villenas2025ofc} and the three constellations from Fig.~\ref{fig:constellations}. To simulate different received optical powers, we varied the OMA of the signal. We choose three different RIN values of $-147$, $-144$, and $-141$ dB/Hz. Figure~\ref{fig:SER} shows the resulting SNR, SER, and BER.

The SNR results are shown in Fig.~\ref{fig:SER}(a), where SNR is calculated analytically using \eqref{eq:snr}. 
We observe that the SNR saturates as OMA increases due to RIN, as previously shown in Ch.~10.6.3 of \cite{hui2019introduction}. For lower RIN values, the SNR saturates at a higher value. Both $\mathcal{X}_{32}^{\mathrm{ref}}$ and $\mathcal{X}_{32}^{\star}$ achieve gains in SNR over $\mathcal{X}_{32}^{+}$. 
After saturation, our optimized constellation improves the SNR by 0.94~dB over $\mathcal{X}_{32}^+$, while $\mathcal{X}_{32}^{\mathrm{ref}}$ achieves a gain of 0.71~dB, as shown in the inset of Fig.~\ref{fig:SER}(a). From \eqref{eq:snr_inf} in dB units, we have that the SNR has a linear dependence with the RIN parameter. Thus, $\mathcal{X}_{32}^\star$ is expected to achieve, with higher RIN, a similar SNR to $\mathcal{X}_{32}^+$.
This is observed in the inset of Fig.~\ref{fig:SER}(a), where $\mathcal{X}_{32}^\star$ with $-144$ dB/Hz RIN almost reaches the saturation SNR of a system with $\mathcal{X}_{32}^+$ and $-145$ dB/Hz RIN. 

To obtain the SER and BER results shown in Fig.~\ref{fig:SER}(b-c), we employed Monte Carlo simulations and the decision rule in \eqref{eq:ml}.
For all considered RIN values, both SER and BER have a flooring behavior as the OMA increases as a result of the SNR saturation. In this saturation regime, $\mathcal{X}_{32}^+$ gives the highest error rates, while $\mathcal{X}_{32}^{\mathrm{ref}}$ only slightly improves SER and BER. As shown in the insets of Fig.~\ref{fig:SER}(b-c), $\mathcal{X}_{32}^\star$ shows visible gains and achieves a greater reduction in SER and BER in the high OMA regime.



Fig.~\ref{fig:SER}(c) also shows that the proposed labeling in Fig.~\ref{fig:constellations}(c) provides BER behavior similar to a Gray labeling at asymptotically high OMAs, i.e., the BER approaches $\mathrm{SER}/\log_2(32)$ (horizontal lines in Fig.~\ref{fig:SER}(c))~\cite{agrell2006gray}.
Therefore, we conjecture that the proposed labeling is the optimal labeling for our optimized QAM-32 in the error floor region, and any improvements by jointly optimizing the constellation $\mathcal{X}_{32}$ and the labeling will be negligible.



\section{Conclusions}

We proposed a novel 2D modulation format, based on PAM-6, optimized given the RIN limitation for next-generation intra-DC interconnect links. Compared to conventional QAM-32, our proposed format achieves an SNR gain of 0.94~dB, and offers the lowest SER from a 32-point 2D constellation generated from 1D PAM-6. 
We also heuristically designed a binary labeling that minimizes the BER for this new constellation and is asymptotically optimal, given that it achieves performance close to that of a Gray labeling in the error floor region.
Future work will focus on extending the error probability analysis of this new constellation and also include geometric shaping.



\section{Acknowledgements}
This research is part of the project COmplexity-COnstrained LIght-coherent optical links \mbox{(COCOLI)} funded by Holland High Tech $|$ TKI HSTM via the PPS allowance scheme for public-private partnerships.

\printbibliography[]



    

\vspace{-4mm}

\end{document}